\documentclass[aps,showpacs,superscriptaddress]{revtex4}
\usepackage{amscd}
\usepackage{amsmath}
\usepackage{amssymb}
\usepackage{amsthm}
\usepackage{graphicx}
\usepackage{mathrsfs}
\usepackage{hyperref}
\usepackage{mathptmx}
\usepackage{subfig}
\usepackage{color}
\usepackage{enumerate}
\usepackage{comment}
\newcommand{\al}{\alpha}
\newcommand{\be}{\beta}

\newcommand{\De}{\Delta}

\newcommand{\fr}{\frac}
\newcommand{\Ga}{\Gamma}
\newcommand{\ga}{\gamma}

\newcommand{\lf}{\left}

\newcommand{\pa}{\partial}

\newcommand{\rg}{\right}

\newcommand{\si}{\sigma}
\newcommand{\te}{\theta}

\newcommand{\GIGa}{\text{GIGa}}

\begin{document}
\title{Distribution Of Wealth In A Network Model Of The Economy}
\author{Tao Ma}
\affiliation{Department of Physics, University of Cincinnati, Cincinnati, OH 45244-0011}
\author{John G. Holden}
\email{john.holden@uc.edu}
\affiliation{CAP Center for Cognition, Action, and Perception, Department of Psychology, University of Cincinnati, Cincinnati, OH 45244-0376}
\author{R.A. Serota}
\email{serota@ucmail.uc.edu}
\affiliation{Department of Physics, University of Cincinnati, Cincinnati, OH 45244-0011}

\date{\today}
\begin{abstract}
We show, analytically and numerically, that wealth distribution in the Bouchaud-M\'{e}zard network model of the economy is described by a three-parameter generalized inverse gamma distribution. In the mean-field limit of a network with any two agents linked, it reduces to the inverse gamma distribution.
\end{abstract}
\maketitle

\section{Introduction}

Wealth distribution has become a subject of keen interest in econophysics research.\cite{yakovenko2009} Here, we study a network model of the economy proposed by Bouchaud and M\'{e}zard (BM).\cite{bouchaud2000} In the mean field (MF) limit of a completely connected network, where any two agents in the network are linked, the model yields the inverse gamma (IGa) stationary wealth distribution. An important feature of the IGa distribution is the power-law (PL) tail.\cite{fujiwara2003} In the opposite limit of a completely disconnected network, the time-dependent part of wealth distribution is lognormal (LN). Both LN and IGa have long history in models of wealth distribution. LN is generated by Gibrat's law.\cite{aoyama2010} IGa, as well as a specific form of GIGa (generalized inverse gamma distribution), were used to analyze wealth distribution in ancient Egypt.\cite{abul2002}

Souma \emph{et al.} did a numerical study of the BM model and proposed that there may be quite an abrupt transition between IGa to LN as a function of the number of connections and the type of connections -- regular network or a small-world network. \cite{souma2001} In this paper, we revisit Souma's simulations and compute the p-values of the fitting distributions using Kolmogorov-Smirnov test \cite{clauset2007}. We argue that the time-dependent LN distribution is a transient -- albeit possibly slow, depending on the parameters -- and concentrate on the stationary solution. We find that for the BM model the latter is a three-parameter GIGa distribution. Theoretically, we develop an effective field theory for the BM model of partially connected networks, including regular network and a random small-world network \cite{souma2001} and obtain the Fokker-Planck equation (FPE) for the probability density function (PDF). Its stationary solution is a GIGa distribution, with IGa distribution as its limit in the MF regime.

This paper is organized as follows. In Sec. \ref{sec:theory}, we discuss the effective field theory of the BM model, the corresponding stationary FPE and its GIGa solution. In Sec. \ref{sec:num}, we present the results of our numerical simulations. In Sec. \ref{sec:sum}, we summarize our findings and outline future directions of our work.

\section{Theory}\label{sec:theory}

\subsection{GIGa from Bouchaud-M\'{e}zard model}

The BM model reads \cite{bouchaud2000}:
\begin{equation}\label{eq:BM_model}
dW_i \overset{S}{=} \sqrt{2} \si W_i dB_i  + \sum_{j(\neq i)} J_{ij} W_j dt - \sum_{j(\neq i)} J_{ji} W_i dt,
\end{equation}
where $\overset{S}{=}$ means that the stochastic differential equation (SDE) is interpreted in the Stratonovich sense \cite{bouchaud2000,souma2001} and $i=1,2,...,N$ with $N \gg 1$ the total number of agents, $W_i$ is the wealth of an agent, $dB_i$ is an independent Wiener process and $\sigma$ and $J_{ij}$ are constants. Since the BM model may have a wider range of applications -- including possibly neural networks -- than originally intended thus we will study it without applying specific interpretations to $W$ and the model parameters.

The BM model in (\ref{eq:BM_model}) can be rewritten into an Ito SDE \cite{jacobs2010}
\begin{equation}\label{eq:Souma}
dW_i \overset{I}{=} \sqrt{2}\si W_i dB_i + \si^2 W_i dt + \sum_{j(\neq i)} J_{ij} W_j dt - \sum_{j(\neq i)} J_{ji} W_i dt .
\end{equation}
in agreement with Souma \emph{et al}. \cite{souma2001}. Rescaling per $W_i(t) =  w_i(t) e^{\si^2 t}$, we obtain
\begin{equation}\label{eq:economic_network_Ito}
dw_i \overset{I}{=} \sqrt{2}\si w_i dB_i + \sum_{j(\neq i)} J_{ij} w_j dt - \sum_{j(\neq i)} J_{ji} w_i dt .
\end{equation}
It is easily seen that in the large $N$ limit, $\sum_{i=1}^N dw_i =\sqrt{2}\si \sum_{i=1}^N w_i dB_i \approx 0$, which implies that the total ``wealth'' fluctuates around a constant value.

Ultimately, the goal is to determine the PDF $P(w,t)$. Towards this end we notice that there is a discontinuous transition from the interacting case $J_{ij} \ne 0$ to the non-interacting case $J_{ij} = 0 $, that is, as soon as the interaction between the agents is turned on, the nature of the distribution function is qualitatively changed. Indeed, as follows from eq. (7.8) in \cite{jacobs2010}, $P(w,t)$ does not have a stationary limit for $J_{ij}=0$ and decreases to zero for any finite $w$ when $t\rightarrow +\infty$ (while preserving the total ``wealth''):
\begin{equation}\label{eq:J0}
P(w,t) = \fr{1}{2\sqrt{\pi t} \si w} \exp\lf[ -  \fr{1}{2} \lf(\fr{\log w + \si^2 t}{\sqrt{2t}\si}\rg)^2 \rg].
\end{equation}
Conversely, in the $J_{ij} \neq 0$, a stationary solution $P(w) \equiv P(w,\infty)$ exists and in what follows we concentrate on its analytical derivation while leaving dynamics to numerical investigation.

The MF limit of a completely connected network was studied in \cite{bouchaud2000}. Substituting $J_{ij}=J/N$ in (\ref{eq:economic_network_Ito}) and extending summation on $j$ to each member of the network, we obtain
\begin{equation}\label{eq:economic_network_mft}
dw_i \overset{I}{=} \sqrt{2} \si w_i dB_i + J (\overline{w} - w_i) dt,
\end{equation}
where $\overline{w}=N^{-1}\sum_{i=1}^N w_i$ is the average of $w_i$. The corresponding FPE is given by
\begin{equation}\label{eq:FokkerPlanck_mft}
\fr{\pa P(w)}{\pa t} = \fr{\pa[ J(w-\overline{w}) + \si^2 w] P}{\pa w} + \si^2 \fr{\pa}{\pa w} \lf[ w \fr{\pa w P}{\pa w} \rg].
\end{equation}
Rescaling via $w \rightarrow w/\overline{w}$, so that $\overline{w} = 1$, we find the normalized stationary IGa solution \cite{bouchaud2000}
\begin{equation}\label{eq:P_mft}
P(w) =  \fr{ \lf(\fr{  \si ^2}{ J}\rg)^{-\fr{J+\si ^2}{  \si^2}} }{\Ga \lf(\fr{J+\si ^2}{  \si ^2}\rg) }
   e^{-\fr{J }{\si^2 } w^{-1}} w^{-2-\fr{J}{\si^2}} .
\end{equation}
with a PL tail $\propto w^{-(2+J/\si^2)}$.

For a partially connected network, where each agent is connected with $1 \le n = zN \le(N-1)$ other agents ($0 < z < 1$), we substitute $J_{ij}=J/n$ in (\ref{eq:economic_network_Ito}) and notice that
\begin{equation}\label{eq:eq:economic_network_pcn}
\sum_{\text{interacting agents: }j(\neq i)} J_{ij} (w_j - w_i)= J (\overline{w}^{(n)} - w_i),
\end{equation}
where $\overline{w}^{(n)}=n^{-1}\sum_{\text{interacting agents: }j(\neq i)} w_j$ is the average over interacting agents. Observing that $\overline{w}^{(n)} \sim w_i$ when $n \sim 1$ and  $\overline{w}^{(n)} \sim 1$ when $n \sim N$, we introduce an \emph{effective field theory ansatz}:
\begin{equation}\label{ansatz}
\overline{w}^{(n)} \rightarrow \te(\ga) w^{1-\ga},
\end{equation}
where $\ga=1$ corresponds to the MF limit and $\ga \rightarrow 0$ to the minimally connected network. The corresponding FPE becomes
\begin{equation}\label{eq:FokkerPlanck}
\fr{\pa P(w)}{\pa t} = \fr{\pa[ J(w-\te w^{1-\ga}) + \si^2 w] P}{\pa w} + \si^2 \fr{\pa}{\pa w} \lf[ w \fr{\pa w P}{\pa w} \rg]
\end{equation}
which has the normalized GIGa solution
\begin{equation}\label{eq:P}
P(w) =  \fr{ \ga  \lf(\fr{\ga  \si ^2}{\te J}\rg)^{-\fr{J+\si ^2}{\ga  \si^2}} }{\Ga \lf(\fr{J+\si ^2}{\ga  \si ^2}\rg) }
   e^{-\fr{J\te }{\si^2 \ga} w^{-\ga}} w^{-2-\fr{J}{\si^2}} .
\end{equation}
with the same PL tail $\propto w^{-(2+J/\si^2)}$ as we find in the MF limit. $\te(\ga)$ is determined from the normalization condition
\begin{equation}\label{eq:wmean}
\overline{w} = \int_0^{\infty}\! w P(w) \, \mathrm{d} w = 1,
\end{equation}
and is given by
\begin{equation}\label{eq:theta}
\te(\ga) = \frac{\gamma  \sigma ^2}{J}\left(\frac{\text{Gamma}\left[\frac{J+\sigma ^2}{\gamma  \sigma ^2}\right]}{\text{Gamma}\left[\frac{J}{\gamma  \sigma ^2}\right]}\right)^{\gamma }.
\end{equation}
It is a monotonic functions between the endpoints $\te(1)=1$ and $\te(0)=1+\si ^2/J$ respectively.

We emphasize that (\ref{eq:P}) describes a GIGa distribution for any $\ga \in (0,1]$ with the limit of IGa for $\ga=1$. $\ga$ has a finite lower cut-off even if only a few connections for each agent are present, while $\ga=0$ corresponds to a completely disconnected network which does not have a stationary solution. The latter was explained before but also follows from (\ref{eq:FokkerPlanck}) since the term in parentheses (multiplying $J$) in the r.h.s. is zero in the $\ga=0$ limit.

\section{Numerical simulation of Bouchaud-M\'{e}zard model}\label{sec:num}

We employ the numerical algorithm described in the Appendix. The time evolution of the distribution function (and its parameters) is observed on approach to its stationary limit. Two cases of the network model are considered: \cite{souma2001}
\begin{enumerate}
\item In a regular network (RegN), each agent connects with $1 \le n=zN \le(N-1)$ nearest neighbors on a circle. In the simplified model, we set $J_{ij} = J/n$ in the numerical simulation.
\item In a random small-world network (RanN), any two agents on a circle have a probability $p_{\text{SW}}$ to be connected. In the simplified model, we set $J_{ij} = J/n$ in the numerical simulation, where $n=p_{\text{SW}}N$. \footnote{We use the symbol $p_{\text{SW}}$ to be consistent with \cite{souma2001}.}
\end{enumerate}
Not surprisingly, the main differences between the two networks are as follows:
\begin{itemize}
\item RanN has considerably shorter equilibration time than RegN on approach to a stationary distribution;
\item RanN is better described by EFT than RegN.
\end{itemize}

The GIGa PDF is given by
\begin{equation}\label{eq:GIGa}
\GIGa(\al, \be, \ga; w)
= \fr{\ga}{\be\Ga(\al )} e^{-\lf(\fr{\be }{w}\rg)^\ga} \lf(\fr{\be }{w}\rg)^{1+\al\ga}.
\end{equation}
While in theoretical description above the mean is set to unity (\ref{eq:wmean}), which would stipulate $\be=\Ga(\al)/\Ga(\al-1/\ga)$, numerically we fit with a three-parameter GIGa, which allows deviation from $\overline{w} = 1$. In our simulation we use $\si = \sqrt{0.05}$ and $J=0.1$. Comparison of (\ref{eq:GIGa}) with (\ref{eq:P}) yields

\begin{equation}\label{eq:params}
\al \ga = 1+J/\si^2=3,
\end{equation}
whence in the MF theory limit, $\ga=1$, we have $\al=3$ and $\be=2$.

\begin{figure}[htp]
\centering
\includegraphics[width=0.35\textwidth]{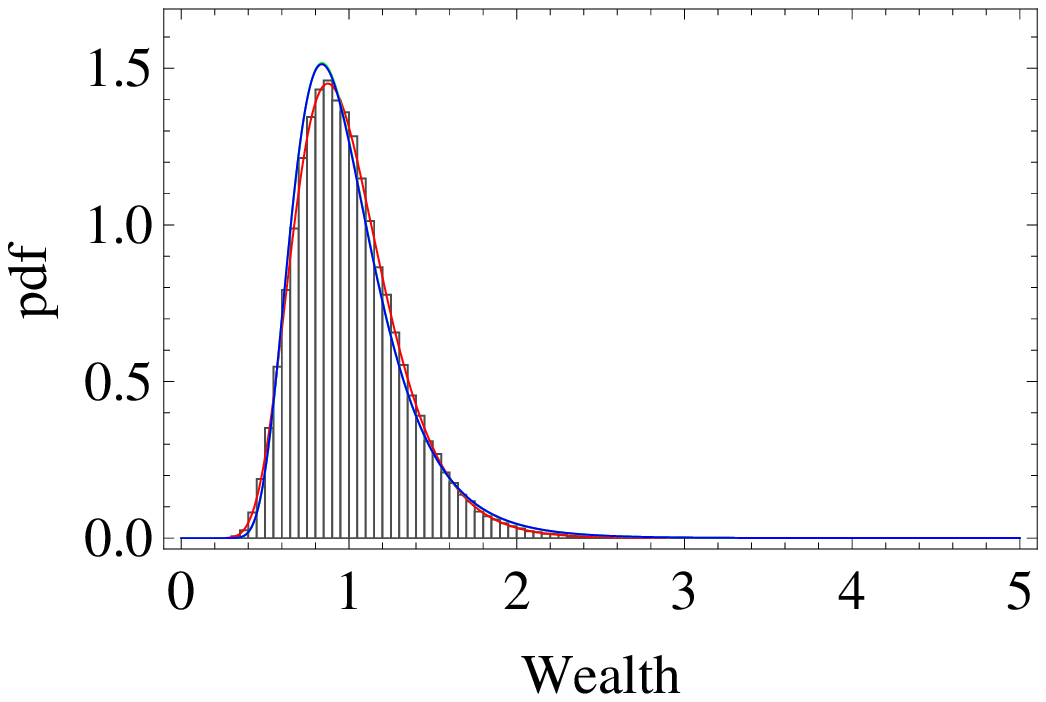}
\includegraphics[width=0.35\textwidth]{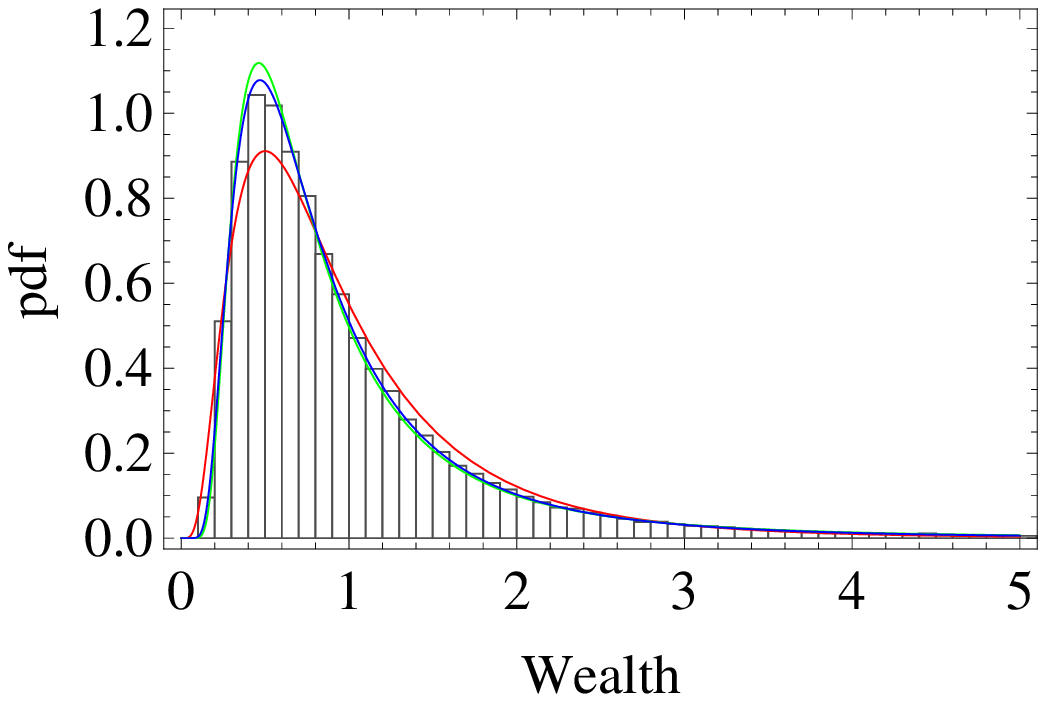}
\caption{Histogram of ``wealth'' distribution for regular network with $z=0.01$ at $t=1$ (left) and $t=2500$ (right). Red: LN; Green: IGa; Blue: GIGa. The fitting of LN is good only for very short times. At $t=1$, the p-value of LN is 0.18; of IGa is 0; of GIGa is 0. At $t=2500$, the p-value of LN is 0; of IGa is 0; of GIGa is 0.05. For $t\ge 2$, the p-value of LN is always 0.}
\label{fig:histogram_RegN}
\end{figure}

\begin{figure}[htp]
\centering
\includegraphics[width=0.35\textwidth]{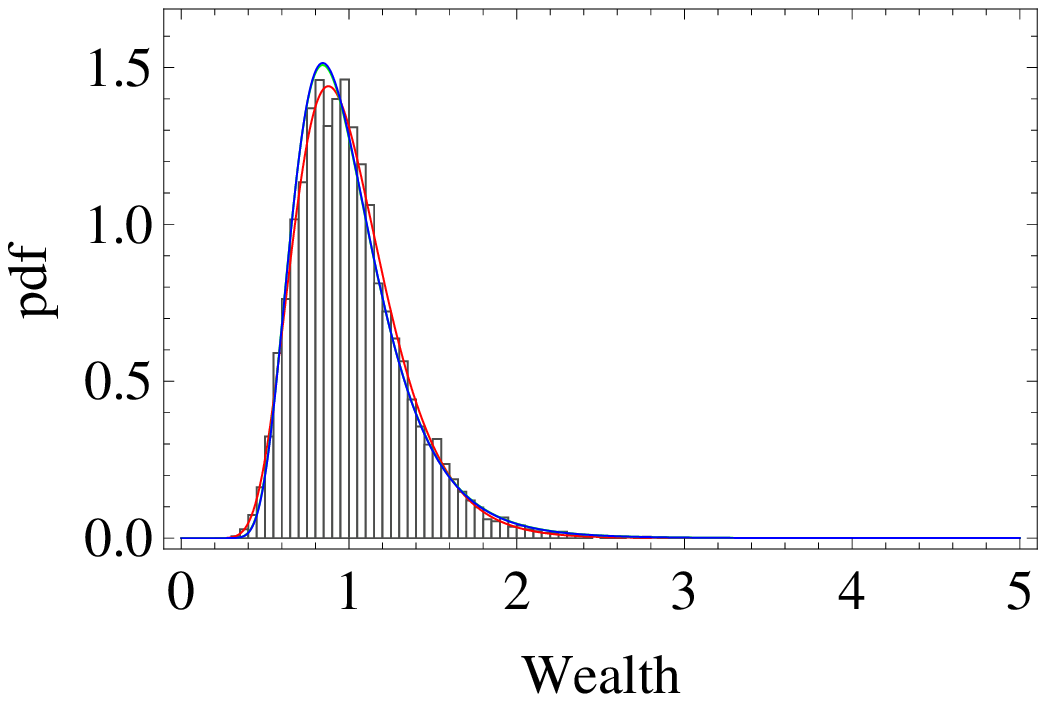}
\includegraphics[width=0.35\textwidth]{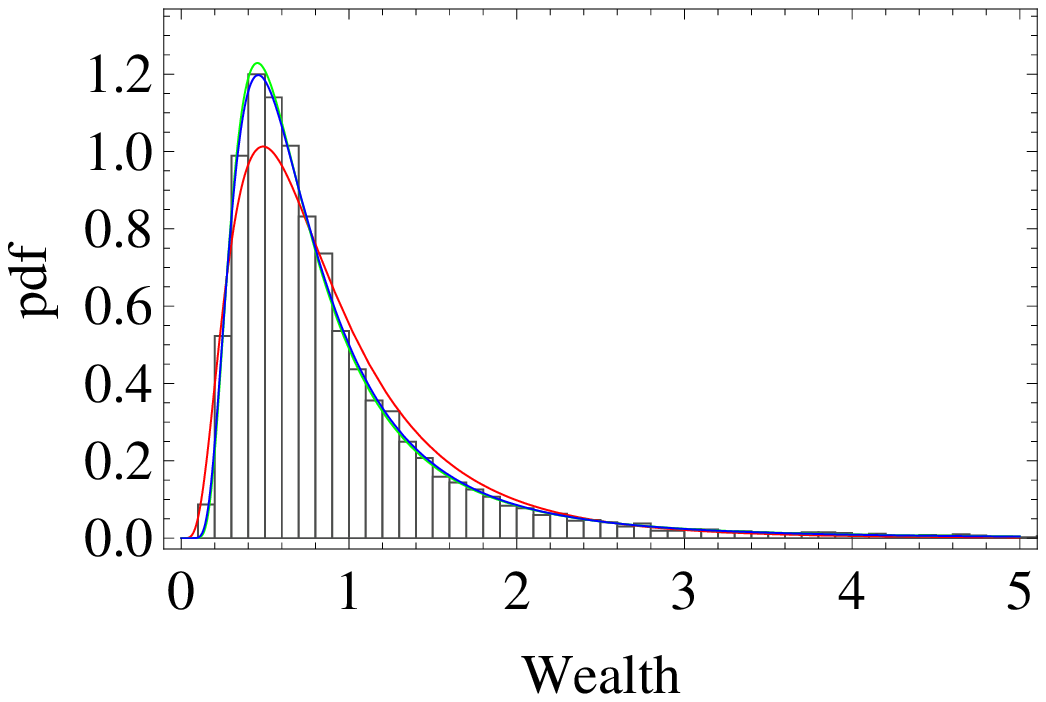}
\caption{Histogram of ``wealth'' distribution for random small world network with $p_{SW}=0.003$ at $t=1$ (left) and $t=500$ (right). Red: LN; Green: IGa; Blue: GIGa. The fitting of LN is good only for very short times. At $t=1$, the p-value of LN is 0.51; of IGa is 0; of GIGa is 0. At $t=500$, the p-value of LN is 0; of IGa is 0.006; of GIGa is 0.97. For $t\ge 3$, the p-value of LN is always 0.}
\label{fig:histogram_RN}
\end{figure}

\begin{figure}[htp]
\centering
\includegraphics[width=0.35\textwidth]{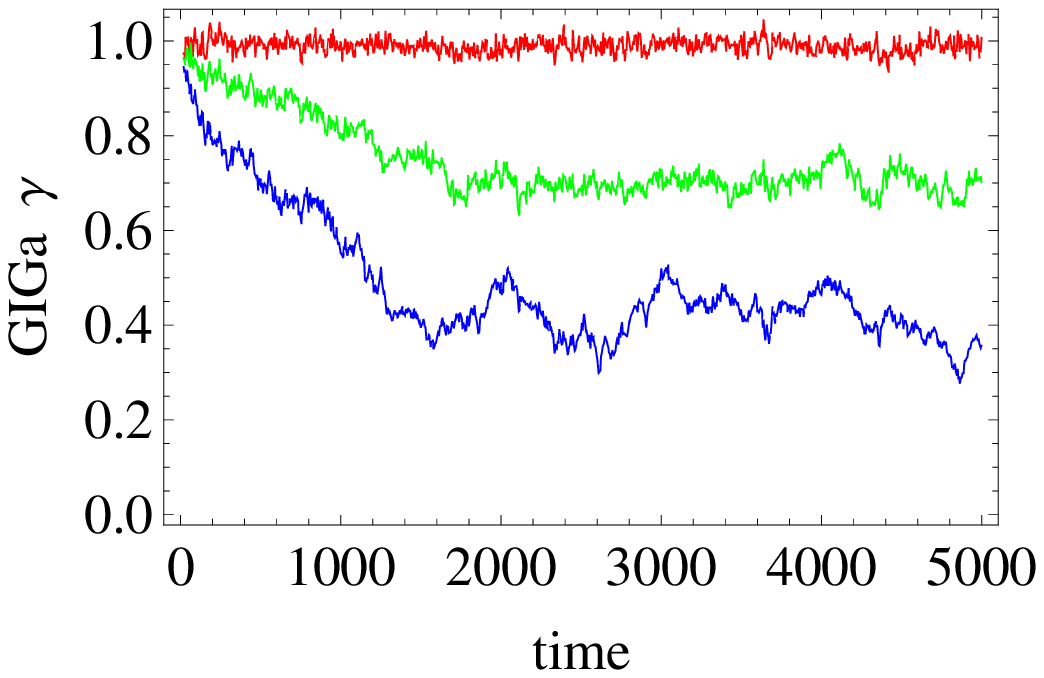}
\includegraphics[width=0.35\textwidth]{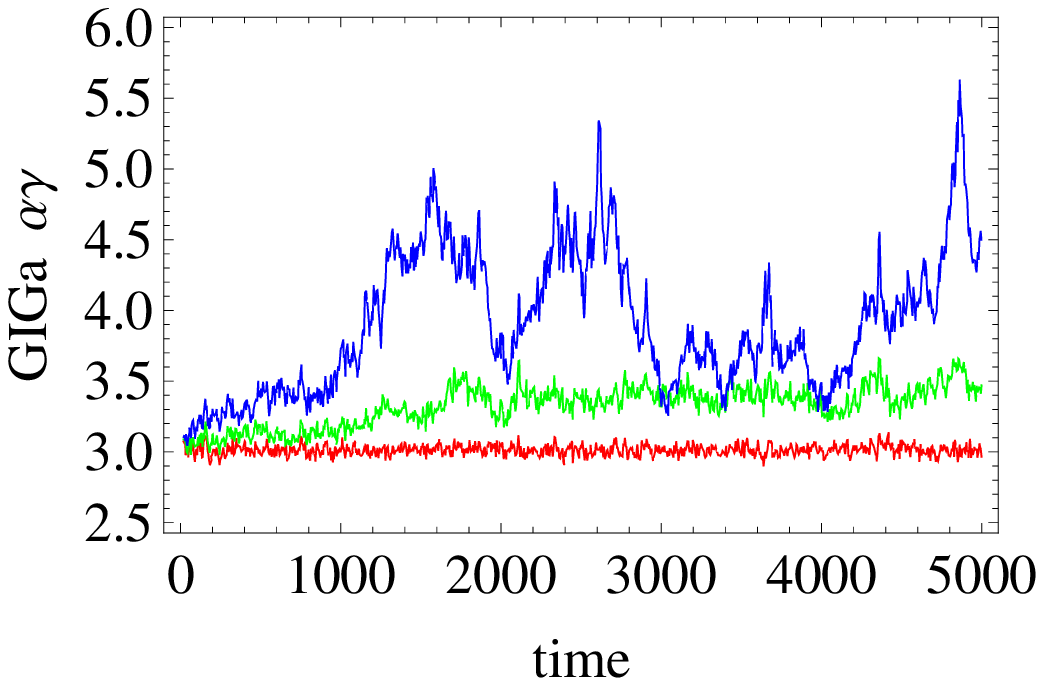}\\
\includegraphics[width=0.65\textwidth]{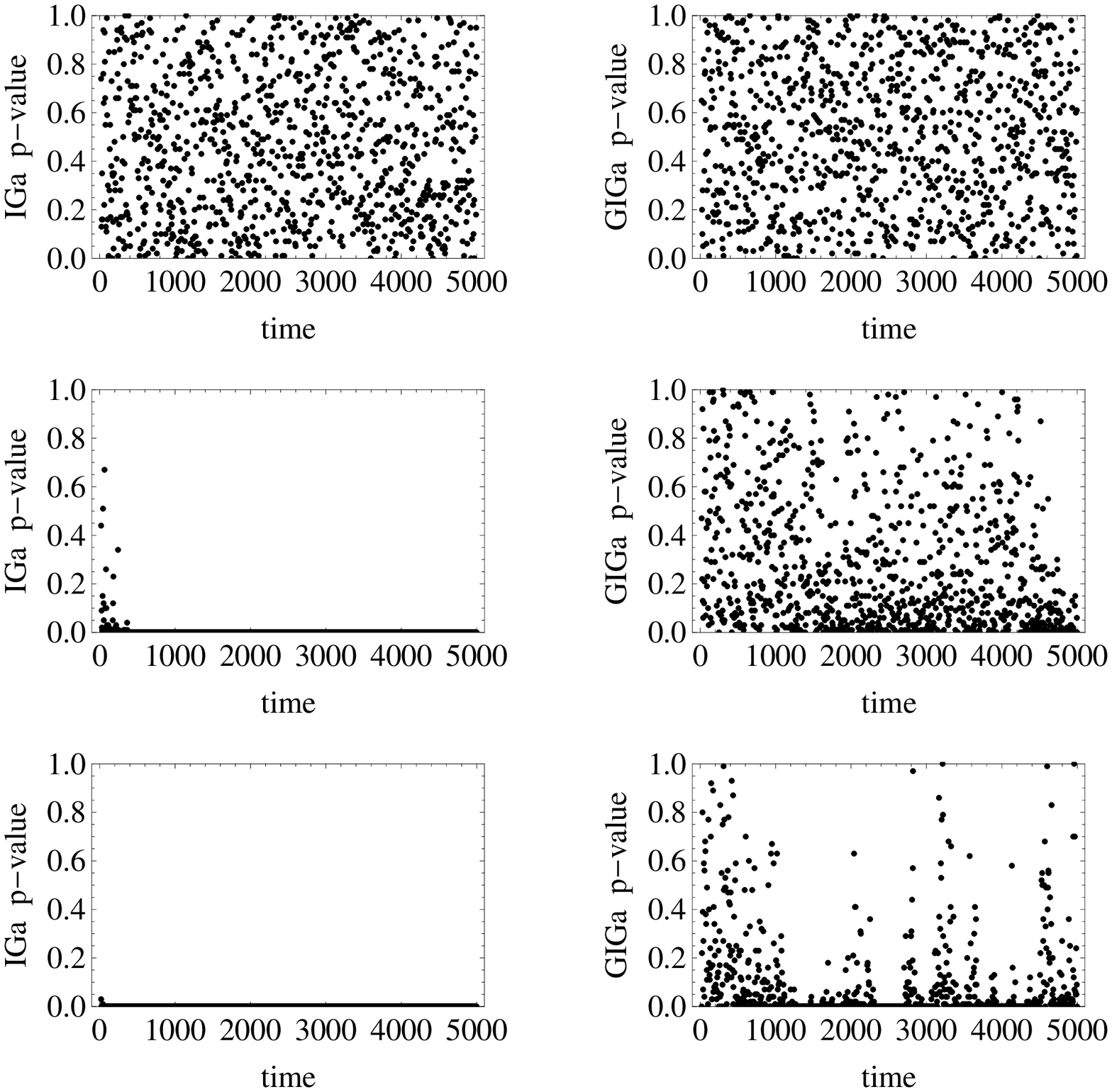}\\
\caption{Regular network. First row: $\ga$ of GIGa (colored), $\al\ga$ of GIGa (colored). Red, green, and blue for $z=0.1, 0.01, 0.003$ respectively. Second, third, and fourth rows (uncolored) are p-values for $z=0.1, 0.01, 0.003$ respectively. }
\label{fig:RegN}
\end{figure}

\begin{figure}[htp]
\centering
\includegraphics[width=0.35\textwidth]{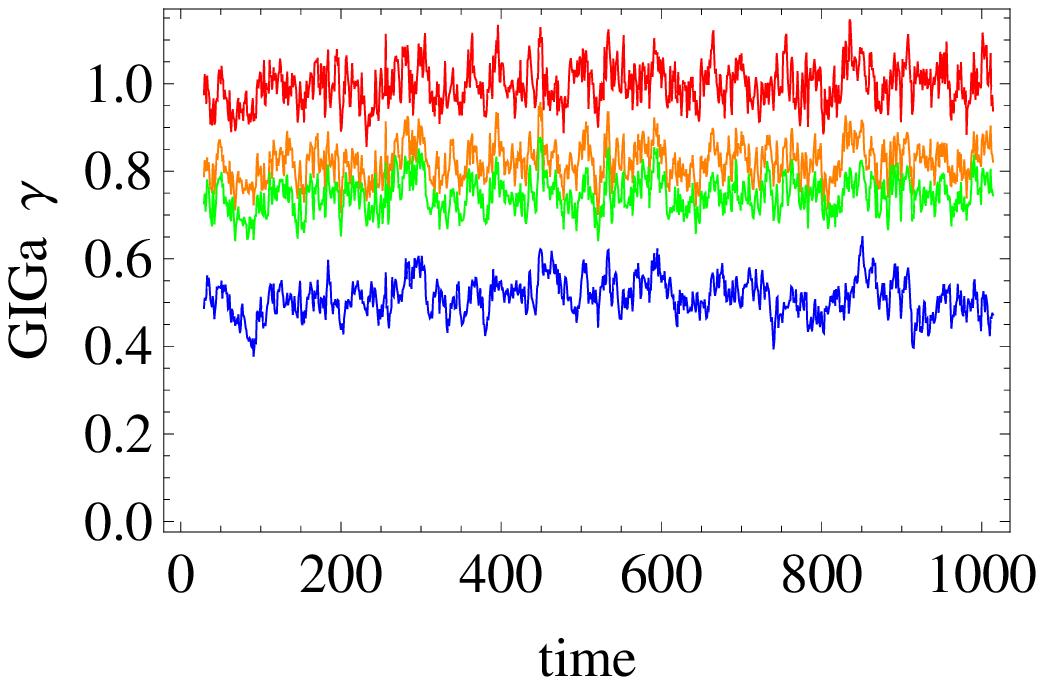}
\includegraphics[width=0.35\textwidth]{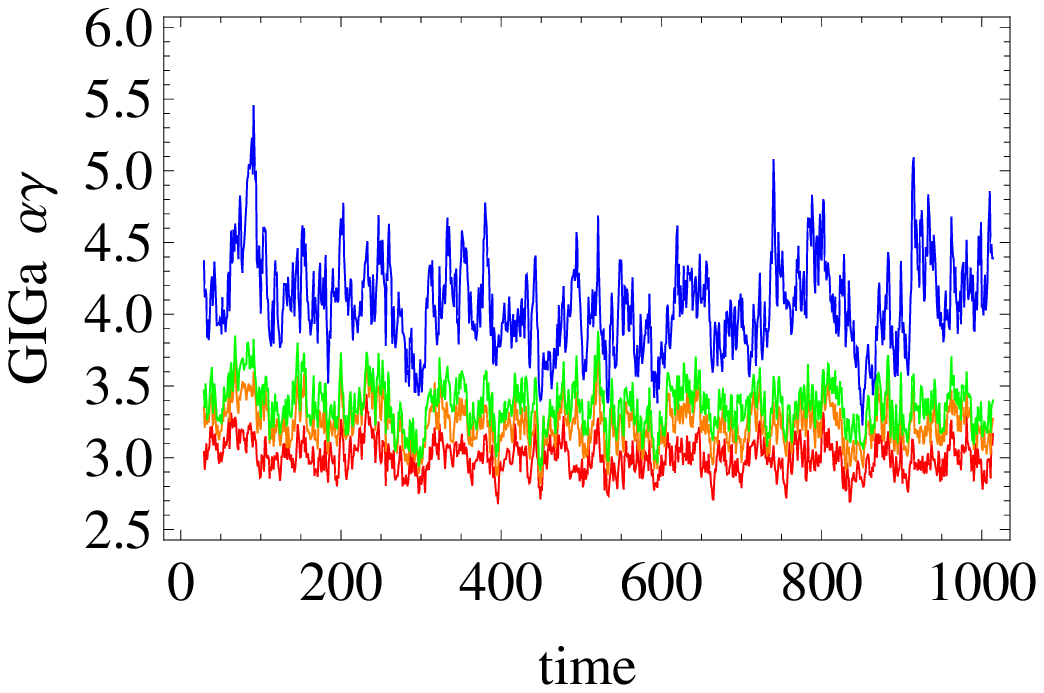}\\
\includegraphics[width=0.65\textwidth]{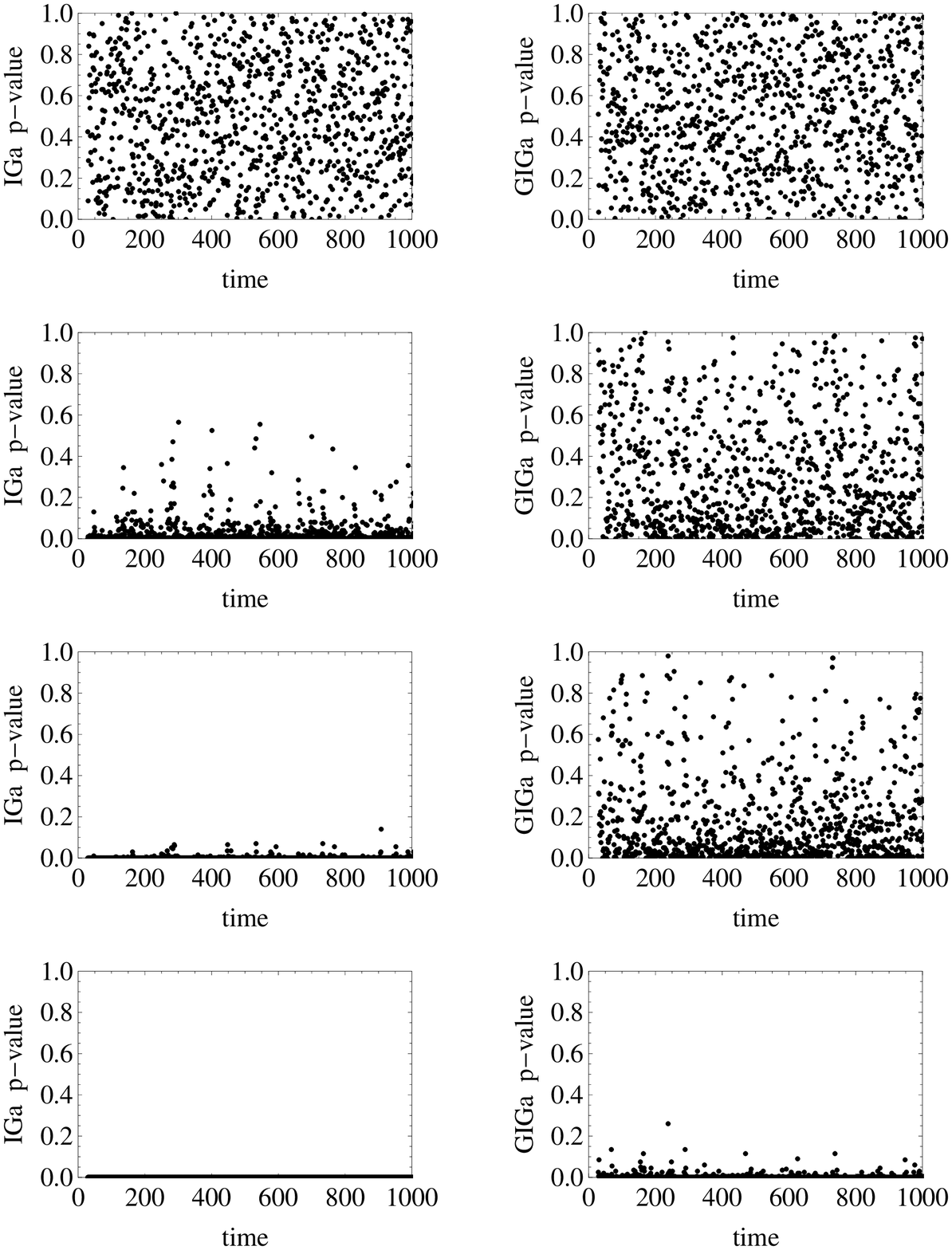}\\
\caption{Random small world network. First row: $\ga$ of GIGa (colored), $\al\ga$ of GIGa (colored). Red, orange, green, and blue for $p_{SW} = 0.1, 0.003, 0.002, 0.001$ respectively. Second, third, fourth and fifth rows (uncolored) are p-values for $p_{SW} = 0.1, 0.003, 0.002, 0.001$} respectively.
\label{fig:RN}
\end{figure}

\begin{figure}[htp]
\centering
\includegraphics[width=0.35\textwidth]{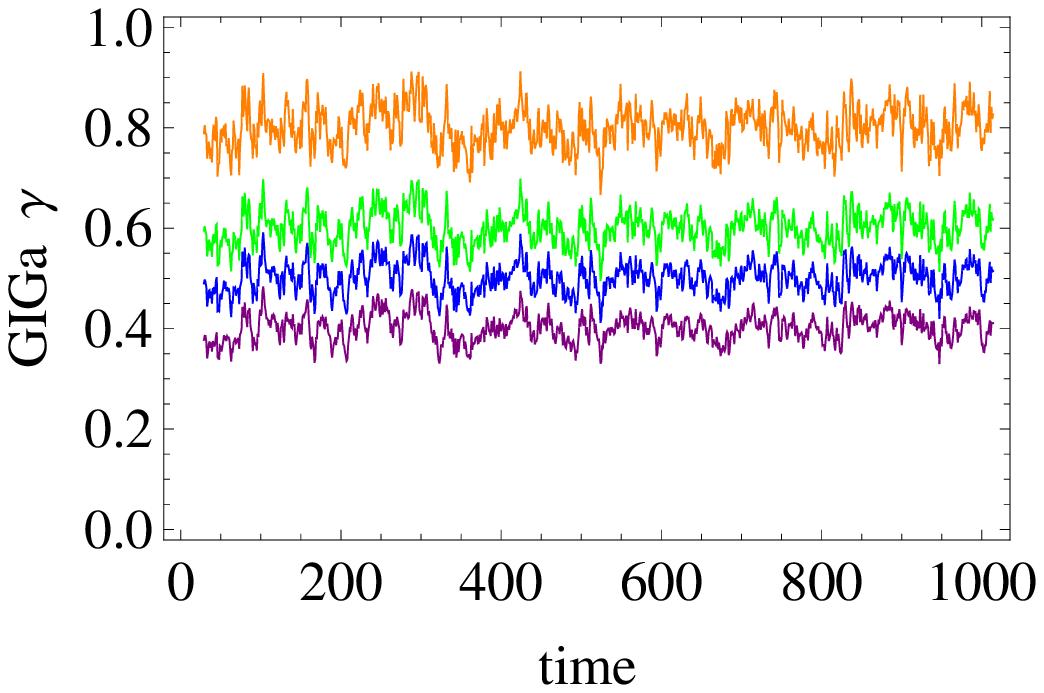}
\includegraphics[width=0.35\textwidth]{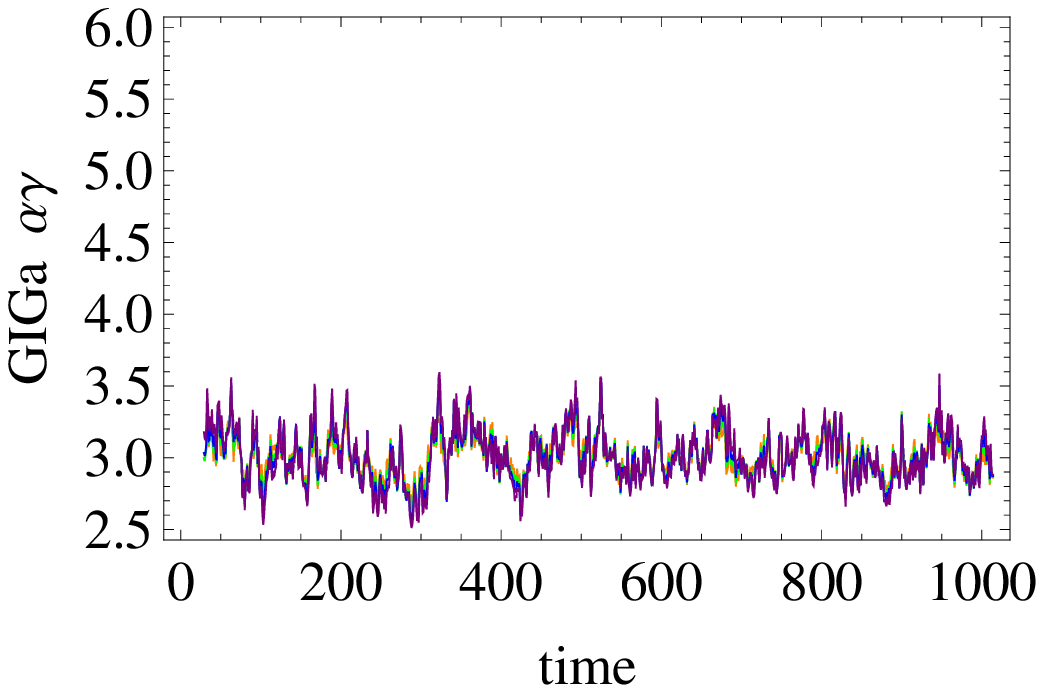}\\
\includegraphics[width=0.65\textwidth]{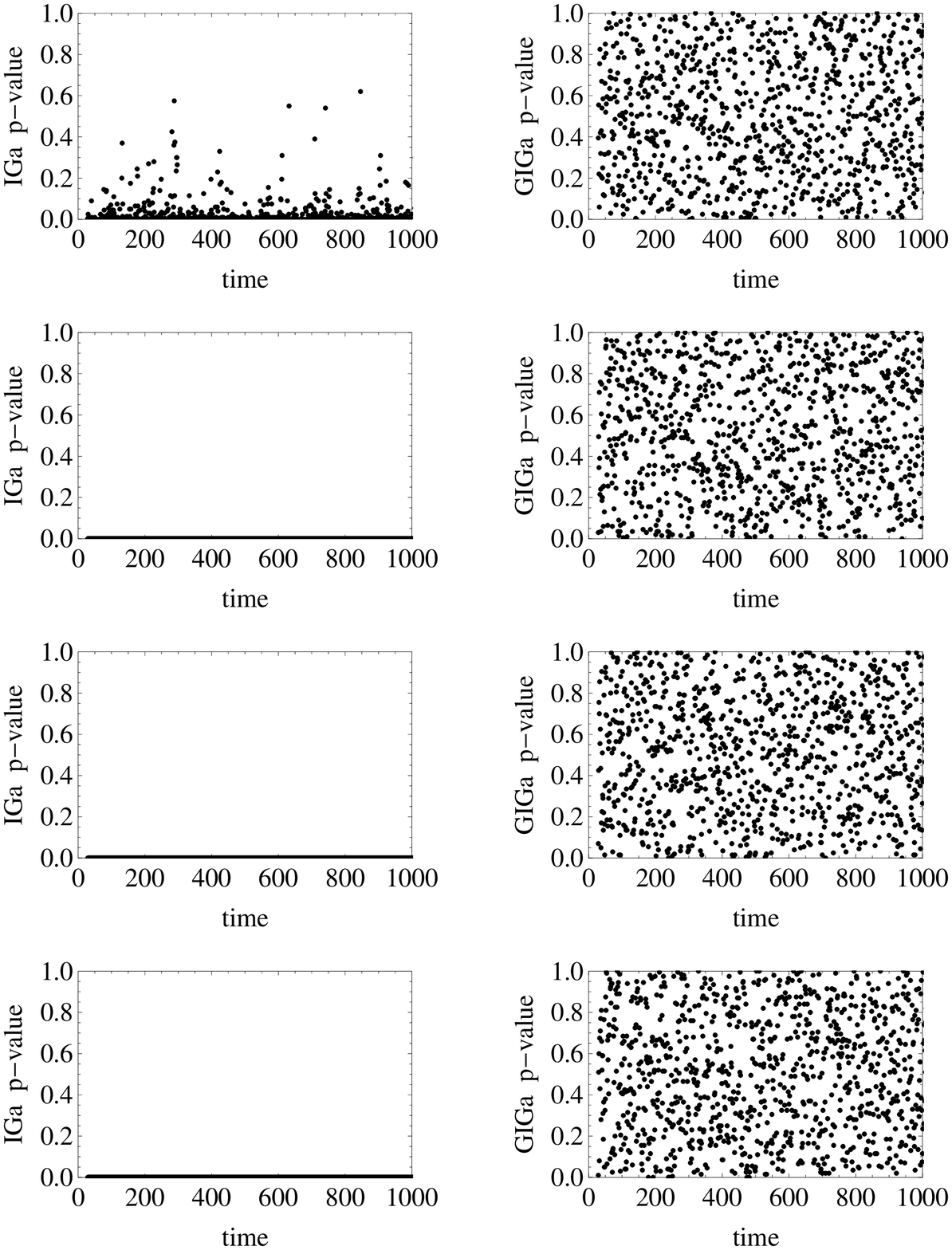}\\
\caption{Effective field theory. First row: $\ga$ of GIGa (colored), $\al\ga$ of GIGa (colored). Orange, green, blue, and purple for $\gamma_{EFT} = 0.8, 0.6, 0.5, 0.4$ respectively. Second, third, fourth and fifth rows (uncolored) are p-values for $\gamma_{EFT} = 0.8, 0.6, 0.5, 0.4$ respectively.}
\label{fig:EFT}
\end{figure}

The results of our numerical simulations are presented as follows. Distribution fitting for RegN and RanN are shown respectively in Figs. \ref{fig:histogram_RegN} and \ref{fig:histogram_RN}. Time evolution of the parameters for RegN and RanN are shown respectively in Figs. \ref{fig:RegN} and \ref{fig:RN}. Fig. \ref {fig:EFT} shows the time evolution of the parameters in the EFT and is color coded to be congruent with Fig. \ref{fig:RN}.

It should be mentioned that we used both the p-value and the log-likelihood to measure which fitting distribution is better, but since they yield the same results, in Figs. \ref{fig:RegN}-\ref {fig:EFT} we present only the former. Also, since the LN fit fails outside initial short time scales for parameters used here, we omit it from these plots as well. Figs. \ref{fig:RegN} and \ref{fig:RN} clearly demonstrate that GIGa provides a better fit than IGa except on approach to MF limit. It is also clear that the equilibration time for establishment of parameters $\ga$ and $\al \ga$ is much faster for RanN than for RegN.\footnote{The initial condition in our simulations was chosen either $w_i(t=0)=1$ or from a narrow Gaussian distribution of $w_i$ centered around 1. Since in the MFT $\overline{w}^{(n)}=\overline{w}$, such choice of initial conditions favors IGa at short times. In Fig.\ref{fig:RegN}, one observes slow equilibration of the RegN from $\gamma=1$ of IGa to $\gamma \le 1$ of the respective GIGa distributions.} Comparison between Figs. \ref{fig:RN} and \ref {fig:EFT} shows that the EFT describes RanN very well.

\onecolumngrid

\section{Discussion}\label{sec:sum}

We demonstrated that the stationary solution of the partially connected Bouchaud-M\'{e}zard network model of the economy predicts the generalized inverse gamma distribution of ``wealth.'' In the mean-field limit of a fully connected network, we recover the inverse gamma distribution obtained in \cite{bouchaud2000}. \footnote{It should be mentioned that the GIGa and IGa are members of a family of distributions related by parametric transformations.\cite{weibullcom,lawless1982,evans2000,watts1998,Crooks2010} For instance, when $\al$ tends to $+\infty$ quadratically and $\ga$ tends to $0^+$ linearly, GIGa tends to a LN distribution. This limiting circumstance corresponds to $J/\sigma^2\to+\infty$. Thus, by tuning parameters $J$ and $\sigma$, the BM model may be able to generate a stationary LN distribution.} Interestingly, the generalized inverse gamma also describes well the distribution of human response times \cite{ma2012}. We speculate that for some cases where Pareto distribution is observed it could be, in fact, a power-law tail of a (generalized) inverse gamma distribution. For instance, the distribution function of landslide area in \cite{turcotte2004}.

We described partially connected networks using an effective field theory and showed that it describes a randomly connected small world network particularly well, including the transitory behavior. Applicability of random versus regular network may depend on the circumstance, given the wide relevance of BM model, ranging from psychology to economics.

An interesting extension of this work would be to study the Watts and Strogatz network model \cite{watts1998}, which has both clustering and small-worldness properties. (This model has already been discussed by Souma \emph{et al.} \cite{souma2001}.) The Watts and Strogatz model is more realistic than RanN as any real economic network should reflect both clustering and small-worldness nature.

\appendix

\section{Numerical simulation method}\label{algorithm}

A comprehensive list of numerical simulation methods of SDEs is given in \cite{kloeden1992}. In this paper, RanN and EFT are simulated by Milstien's method, which has one-order accuracy, and RegN is simulated by the order 1.5 strong Taylor scheme. Below we only present Milstien's method, which can be found in \cite{jacobs2010} and \cite{kloeden1992}. In Milstien's method, the SDE
\begin{equation}
dx \overset{I}{=} f(x,t) dt + g(x,t) dB,
\end{equation}
is written into the difference equation
\begin{equation}
\De x = f \De t + g \De B + \fr{g}{2}\fr{\pa g}{\pa x} [(\De W)^2 - \De t] .
\end{equation}
For the BM model, $g = \sqrt{2} \si w_i $, hence
\begin{equation}
\fr{g}{2}\fr{\pa g}{\pa w_i} = \si^2 w_i
\end{equation}
and
\begin{equation}
\De x = f \De t + g \De B + \si^2 w_i [(\De B)^2 - \De t] .
\end{equation}

\begin{acknowledgements}
We wish to thank Prof. Dr. Igor Sokolov for many helpful discussions at the early stage of this work.
\end{acknowledgements}

\end{document}